\documentclass[reprint,superscriptaddress,amsmath,amssymb, aps, prl, nobibnotes]{revtex4-2}

\PassOptionsToPackage{dvipsnames}{xcolor}
\usepackage{xcolor}

\usepackage[utf8]{inputenc}
\usepackage[T1]{fontenc}
\usepackage[english]{babel}
\usepackage{float}
\usepackage{amsmath}
\usepackage{amsfonts}
\usepackage{amssymb}

\usepackage{graphicx}% Include figure files
\usepackage{dcolumn}% Align table columns on decimal point
\usepackage{bm}% bold math
\usepackage{subfig}
\usepackage{siunitx}
\usepackage{float}
\usepackage{soul}
\usepackage{ulem}
\usepackage[version=4]{mhchem}
\usepackage{xr}
\usepackage{cleveref}
\usepackage{caption}
\usepackage{ragged2e}
\usepackage{float}
\restylefloat{figure}

\usepackage{mathrsfs}

\makeatletter\let\l@en\l@english\makeatother

\begin{document}

\title{Island of Inversion in neutron-rich cobalt isotopes revealed from mass measurements}

\author{M.~Flayol}
\affiliation{Université de Bordeaux, CNRS/IN2P3, LP2I Bordeaux, UMR 5797, F-33170 Gradignan, France}
\author{P.~Ascher}
\email[Contact author: ]{ascher@lp2ib.in2p3.fr}
\affiliation{Université de Bordeaux, CNRS/IN2P3, LP2I Bordeaux, UMR 5797, F-33170 Gradignan, France}
\author{D.D.~Dao}
\affiliation{Université de Strasbourg, CNRS, IPHC UMR 7178, F-67000 Strasbourg, France}
\author{M.~Gerbaux}
\affiliation{Université de Bordeaux, CNRS/IN2P3, LP2I Bordeaux, UMR 5797, F-33170 Gradignan, France}
\author{S.~Grévy}
\affiliation{Université de Bordeaux, CNRS/IN2P3, LP2I Bordeaux, UMR 5797, F-33170 Gradignan, France}
\author{F.~Nowacki}
\affiliation{Université de Strasbourg, CNRS, IPHC UMR 7178, F-67000 Strasbourg, France}
\author{A.~de Roubin}
\affiliation{Université de Bordeaux, CNRS/IN2P3, LP2I Bordeaux, UMR 5797, F-33170 Gradignan, France}
\affiliation{Université de Caen Normandie, ENSICAEN, CNRS/IN2P3, LPC Caen UMR6534, F-14000 Caen, France}
\author{D.~Atanasov}
\affiliation{Université de Bordeaux, CNRS/IN2P3, LP2I Bordeaux, UMR 5797, F-33170 Gradignan, France}
\affiliation{SCK·CEN, Belgian Nuclear Research Centre, Boeretang 200, 2400 Mol, Belgium}
\author{B.~Blank}
\affiliation{Université de Bordeaux, CNRS/IN2P3, LP2I Bordeaux, UMR 5797, F-33170 Gradignan, France}
\author{L.~Canete}
\affiliation{Grand Accélérateur National d’Ions Lourds (GANIL), Bd Henri Becquerel, BP 55027, F-14076 Caen Cedex 5, France}
\affiliation{Normandie Univ., UNICAEN, F-14000 Caen, France}
\author{Q.~Délignac}
\affiliation{Université de Bordeaux, CNRS/IN2P3, LP2I Bordeaux, UMR 5797, F-33170 Gradignan, France}
\author{T.~Eronen}
\author{Z.~Ge}
\affiliation{University of Jyväskylä, Department of Physics, Accelerator Laboratory, P.O. Box 35(JYFL), FI-40014, University of Jyväskylä, Finland}
\author{M.~Hukkanen}
\affiliation{Université de Bordeaux, CNRS/IN2P3, LP2I Bordeaux, UMR 5797, F-33170 Gradignan, France}
\affiliation{University of Jyväskylä, Department of Physics, Accelerator Laboratory, P.O. Box 35(JYFL), FI-40014, University of Jyväskylä, Finland}
\author{A.~Jaries}
\affiliation{University of Jyväskylä, Department of Physics, Accelerator Laboratory, P.O. Box 35(JYFL), FI-40014, University of Jyväskylä, Finland}
\author{A.~Kankainen}
\author{I.D.~Moore}
\author{M.~Mougeot}
\author{A.~Raggio}
\author{J.~Ruotsalainen}
\affiliation{University of Jyväskylä, Department of Physics, Accelerator Laboratory, P.O. Box 35(JYFL), FI-40014, University of Jyväskylä, Finland}
\author{M.~Stryjczyk}
\affiliation{University of Jyväskylä, Department of Physics, Accelerator Laboratory, P.O. Box 35(JYFL), FI-40014, University of Jyväskylä, Finland}
\affiliation{Institut Laue-Langevin, 71 Avenue des Martyrs, F-38042 Grenoble, France}
\author{V.~Virtanen}
\affiliation{University of Jyväskylä, Department of Physics, Accelerator Laboratory, P.O. Box 35(JYFL), FI-40014, University of Jyväskylä, Finland}
\affiliation{Grand Accélérateur National d’Ions Lourds (GANIL), Bd Henri Becquerel, BP 55027, F-14076 Caen Cedex 5, France}

\date{\today}

\begin{abstract}
Mass measurements of the ground and isomeric states of \ce{^{68-70}Co} have been performed using the JYFLTRAP Penning-trap mass spectrometer at the IGISOL facility. 
The masses were measured, either for the first time for the isomeric states of \ce{^{68}Co} and \ce{^{70}Co}, or with greatly improved precision for the others, removing ambiguities in the mass surface beyond $N=40$.
The ordering of the low and high-spin states in \ce{^{68}Co} and \ce{^{70}Co} has also been established. The results, supported by Large-Scale Shell Model and Discrete Non-Orthogonal Shell-Model calculations, show a gradual lowering of intruder states with increasing neutron number, eventually leading to an inversion in \ce{^{70}Co}. These findings clarify previously proposed contradictory interpretations. Finally, we demonstrate the importance of including induced effective 3N forces for a consistent description of binding energies in the island of inversion near $N=40$.
\end{abstract}


\maketitle

\textit{Introduction} - The nuclear mass, directly related to the binding energy, is a fundamental property of atomic nuclei that provides insight into subtle changes in nuclear structure. Binding energies serve as key benchmarks for testing nuclear shell-model predictions, such as the evolution and strength of shell closures at magic neutron ($N$) and proton ($Z$) numbers. Additionally, high-precision mass spectrometry, in particular with the Penning-trap Phase-Imaging Ion-Cyclotron-Resonance (PI-ICR) technique \cite{eliseev_phase-imaging_2013}, enables the measurement of the binding energies of low-energy long-lived isomeric states that are inaccessible to conventional spectroscopy techniques. Excitation energies of isomers provide additional valuable information on nuclear structure and potential shape coexistence and are therefore essential for detailed nuclear structure studies.\\
\indent This work focuses on the $N=40$ region near the semi-doubly magic $^{68}$Ni nucleus, characterized by a closed $Z=28$ proton shell and an $N=40$ neutron sub-shell. Experimental signatures in $^{68}$Ni, such as a relatively high first-excited $2^+$ state energy and a small $B(E2; 0^+ \rightarrow 2^+)$ strength compared to neighboring isotopes~\cite{broda_n40_1995,sorlin__2868ni_40_2002}, support the presence of a sub-shell closure at $N = 40$, although it remains moderate compared to other well-established shell closures. Additionally, the presence of low-lying deformed excited states in $^{68}$Ni  \cite{tsunoda_novel_2014,suchyta_shape_2014} highlights the complexity of this region, located on the shores of the so-called island of inversion (IoI) \cite{thibault_direct_1975, bernas_magic_1982, sorlin_nuclear_2008, lenzi_island_2010, steppenbeck_evidence_2013}.\\
\indent As nucleons are added to or removed from $^{68}$Ni, the nucleus experiences significant changes in its shape and structure. It has been shown in \cite{nowacki_neutron-rich_2021} that removing protons from $^{68}$Ni weakens the neutron $pf$-$gd$ gap. This can be interpreted as the effect of the tensor force between the $\pi f_{7/2}$ and $\nu f_{5/2}$ orbitals \cite{otsuka_novel_2010}. Similarly, adding neutrons in the $g_{9/2}$ orbital reduces the proton gap between the $f_{7/2}$ and $f_{5/2}$ orbitals. These changes facilitate both proton and neutron excitations across the $Z=28$ and $N=40$ gaps, respectively, leading to nucleon configurations with quadrupole correlations\, that in turn favor non-spherical shapes.
The intruder states with such configurations, involving particle-hole ($ph$) excitations, can even become the ground state for cases with the strongest quadrupole correlations, giving rise to the IoI. $^{66}$Fe and $^{64}$Cr, with two and four fewer protons than $^{68}$Ni, respectively, exhibit prolate shapes \cite{rother_enhanced_2011, lunardi_spectroscopy_2007, hannawald_decay_1999, aoi_development_2009, ljungvall_onset_2010}, with the deformation reaching a maximum in $^{64}$Cr, at the center of the IoI. \\
\indent Shape coexistence and intruder states have also been observed in neutron-rich cobalt isotopes, which have only one proton less than Ni. In the odd-even $^{65-69}$Co isotopes, a (1/2$^-$) state, interpreted as a configuration with one proton excitation across $Z=28$, has been observed \cite{liddick_analogous_2015}, with an excitation energy decreasing from \qty[mode=text]{1.095}{MeV} in $^{65}$Co to 182(100) keV 
in \ce{^{69}Co} 
\cite{canete_erratum_2021}. This (1/2$^-$) state appeared to remain above the “spherical” (7/2$^-$) ground state in $^{69}$Co, although its excitation energy in $^{69}$Co was not known precisely.\\
\indent The neutron-rich odd-odd $^{68-70}$Co isotopes also present two long-lived states identified in $\beta$-decay studies \cite{mueller_beta_1999, mueller_ensuremathbeta_2000}: 
a (6$^-$, 7$^-$) state corresponding to the “normal” configuration and a low-spin intruder state exhibiting deformation. For $^{68}$Co and $^{70}$Co, neither the level ordering nor the excitation energies were known. 

The unknown state orderings have led to ambiguities in the mass trend of the Co isotopic chain. 
In particular, Canete et al. \cite{canete_erratum_2021} demonstrated, by measuring the masses of both the ground and isomeric states of $^{69}$Co, that the previous measurement by Izzo et al. \cite{izzo_precision_2018} had mistakenly assigned the isomer to the ground state. Similar misassignments may have occurred in $^{68}$Co and $^{70}$Co, for which only a single state was observed. These ambiguities lead to different trends in the two-neutron separation energy ($S_{2n}$) along the isotopic chain, as shown in Ref.~\cite{canete_erratum_2021}: either a smooth evoultion from $N=39$ to $N=43$, or a sudden change in trend at $N=43$. Precise mass measurements are therefore crucial to clarify the mass surface in this region.

In this context, we report mass measurements of the ground and isomeric states in $^{68-70}$Co. The experimental setup and data analysis are detailed in the following sections, followed by a discussion of the results and their implications for nuclear structure. These data contribute to a refining of the Lenzi-Nowacki-Poves-Sieja (LNPS) interaction \cite{lenzi_island_2010} and a broader study of binding energies in this region.\\

\begin{figure}[h]
\centering
\includegraphics[trim= 0 0 0 0 , clip, width=\columnwidth]{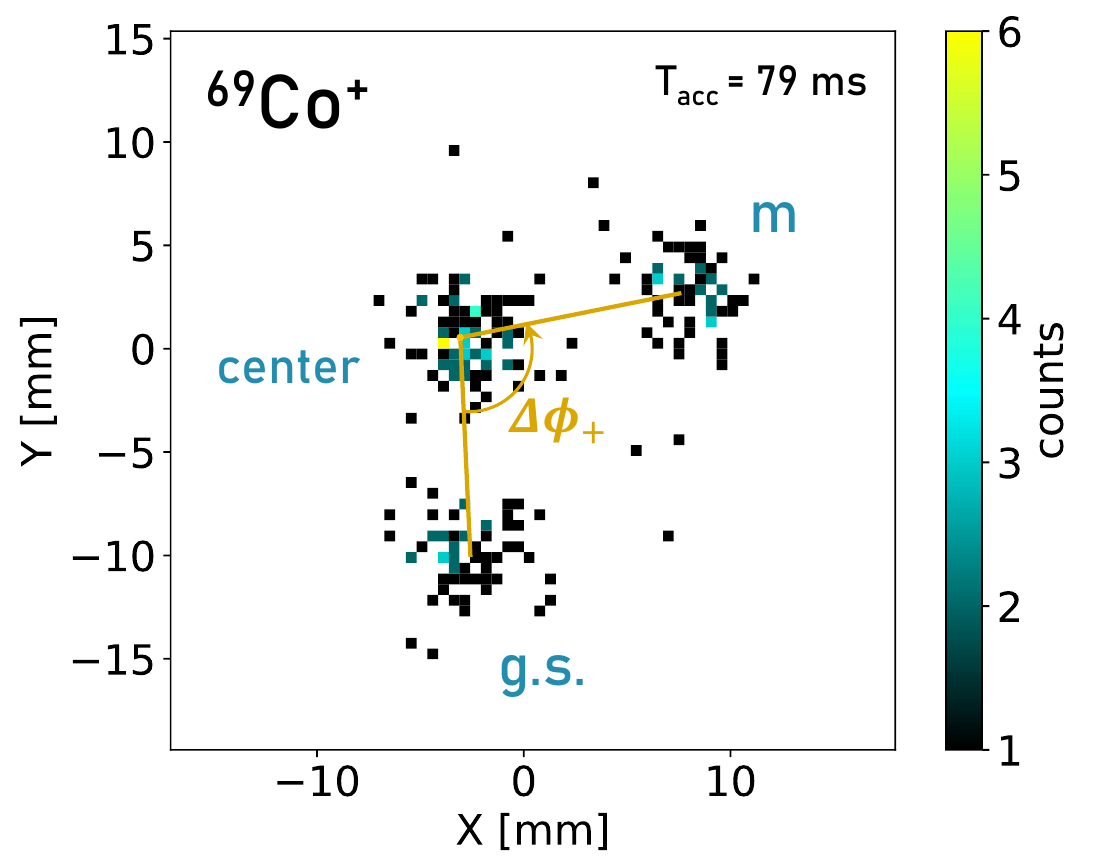}
\caption[]{\justifying \label{fig:co69} Ion projection of $^{69}$Co$^+$ on the position-sensitive detector after magnetron damping only ("center") and after rotation at the reduced cyclotron frequency for T$_{acc}$. The ground state ("g.s.") and isomeric state ("m") are clearly separated in phase by $\Delta \phi_+$.}
\end{figure}

\textit{Results} - The experiment was carried out at the Ion Guide Isotope Separator On-Line (IGISOL) facility~\cite{moore_towards_2013} using the JYFLTRAP double Penning-trap mass spectrometer~\cite{eronen_jyfltrap_2012} at the University of Jyväskylä, Finland. 
Details about the beam production and preparation are given in the Supplemental Material. 
The masses of the ground and isomeric states of $^{68-70}$Co were independently measured for the first time using the PI-ICR technique \cite{eliseev_phase-imaging_2013, nesterenko_phase-imaging_2018}. 
The cyclotron frequency $\nu_c = qB/(2\pi m)$ of an ion with a mass $m$ and a charge $q$ in a magnetic field $B$ 
%($\nu_c\sim$ 1.5 MHz with $B\sim$ 7T) 
was determined from the angular difference between the projections of the radial cyclotron ($\phi_+$) and magnetron ($\phi_-$) phases, accumulated during a time $T_{acc}$, onto a position-sensitive detector (see the time patterns in \cite{nesterenko_study_2021}). Figure~\ref{fig:co69} shows a measurement for $^{69}$Co, where the isomeric state (m) on the right and the ground state (g.s.) at the bottom are clearly separated. The phase difference measured between the two states $\Delta \phi_+$ corresponds to an excitation energy of 148.3(76) keV.
The magnetic field strength $B$ was determined using reference isobaric ions present in the beam. Finally, the atomic mass was derived from the cyclotron frequency ratio $r$ between the reference ion and the ion of interest. Comprehensive details on the systematic aspects of the PI-ICR technique at JYFLTRAP can be found in \cite{nesterenko_phase-imaging_2018,nesterenko_study_2021}. In this work, statistical uncertainties were highly dominant.

Table~\ref{masses} summarizes the results of this work, including the frequency ratios, the mass excesses as well as the spin-parity and half-life orderings of the two long-lived states in $^{68}$Co and $^{70}$Co (presented later in this paper). 

\begin{table*}[hbt!]
\caption[]{\justifying \label{masses} Half-lives ($T_{1/2}$) and spin-parities ($I^{\pi}$) based on NUBASE 2020~\cite{kondev_nubase2020_2021}, mass-excess values from this work (ME$_{exp}$) and from AME 2020 (ME$_{AME}$)~\cite{huang_ame_2021} as well as previous ($E^*_{lit}$) and new ($E^*_{\text{exp}}$) excitation energies for the isomeric states. The values in bold correspond to the cases where the state ordering was newly established in this work. The symbol "\#" denotes a value based on extrapolations. In the last column, the singly-charged ions used to determine the frequency ratios $r=\nu_{ref}/\nu$ are listed. 
\label{tab:massexesses}}
\centering

\scalebox{1}{\begin{tabular}[b]{c|c|c|c|c|c|c|c|c}
    & $T_{1/2}$ [ms]        & $I^{\pi}$   & ME$_{exp}$ [keV]    & ME$_{AME}$ [keV]  & $E^*_{\text{exp}}$ [keV] & $E^*_{\text{lit}}$ [keV]  & $r=\nu_{ref}/\nu$ & Ref              \\ \hline
\rule{0pt}{15pt} $^{68}$Co & \textbf{200(20)}    & $\boldsymbol{(7^-)}$     & $-51652.5(7)$   & $-51643(4)$    &            &           & 1.000290095(8)      & $^{68}$Zn       \\
$^{68m}$Co    & \textbf{1600(300)}  & $\boldsymbol{(2^-)}$     & $-51573.0(21)$  & $-51490(150)\#$ &  79.5(22)  & 150(150)\#  & 1.000291352(32)     & $^{68}$Zn   \\
$^{69}$Co  & 180(20)             & $(7/2^-)$                & $-50320.2(63)$  & $-50390(90)$      &            &           & 0.999997612(72)     & $^{69m}$Co       \\
$^{69m}$Co    & 750(250)            & $1/2^-\#$              & $-50171.9(43)$  & $-50213(13)  $    &  148.3(76) &   170(90)  & 1.000298364(65)     & $^{69}$Ga        \\
$^{70}$Co  & \textbf{508(7)}     & $\boldsymbol{(1^+)}$ & $-46508.1(17)$  & $-46525(11)$      &            &           & 1.000353979(18)     & $^{70}$Zn        \\
$^{70m}$Co    & \textbf{112(7)}     & $\boldsymbol{(7^-)}$ & $-46239.7(47)$  & $-46330(200)\#$ &  268.4(50) & 200(200)\# & 1.000358107(68)     & $^{70}$Zn          \end{tabular}}
\end{table*}

For $^{69}$Co, the mass excesses of both the isomeric and ground states obtained in this work are in agreement with Ref.~\cite{canete_erratum_2021}, confirming that the authors of Ref.~\cite{izzo_precision_2018} likely measured the isomeric state, or a mixture strongly dominated by it. The mass uncertainties have been greatly reduced compared to Ref.~\cite{canete_erratum_2021}, with a 14-fold improvement for the isomeric state.
For $^{68}$Co and $^{70}$Co, the mass excesses of the isomeric states were measured for the first time, and the precision of the ground-state mass excesses were improved. Our results indicate that the mass measurement of $^{70}$Co at JYFLTRAP~\cite{canete_erratum_2021} and $^{68}$Co at LEBIT~\cite{izzo_precision_2018} most likely corresponded to the ground state, or to a mixture dominated by the ground state.
The present mass measurements of pure ground states allows us to clarify the mass surface, that will be described and interpreted in the discussion section.

\begin{figure}[!htb]
%\begin{figure}[H]
\centering
\includegraphics[width=\linewidth]{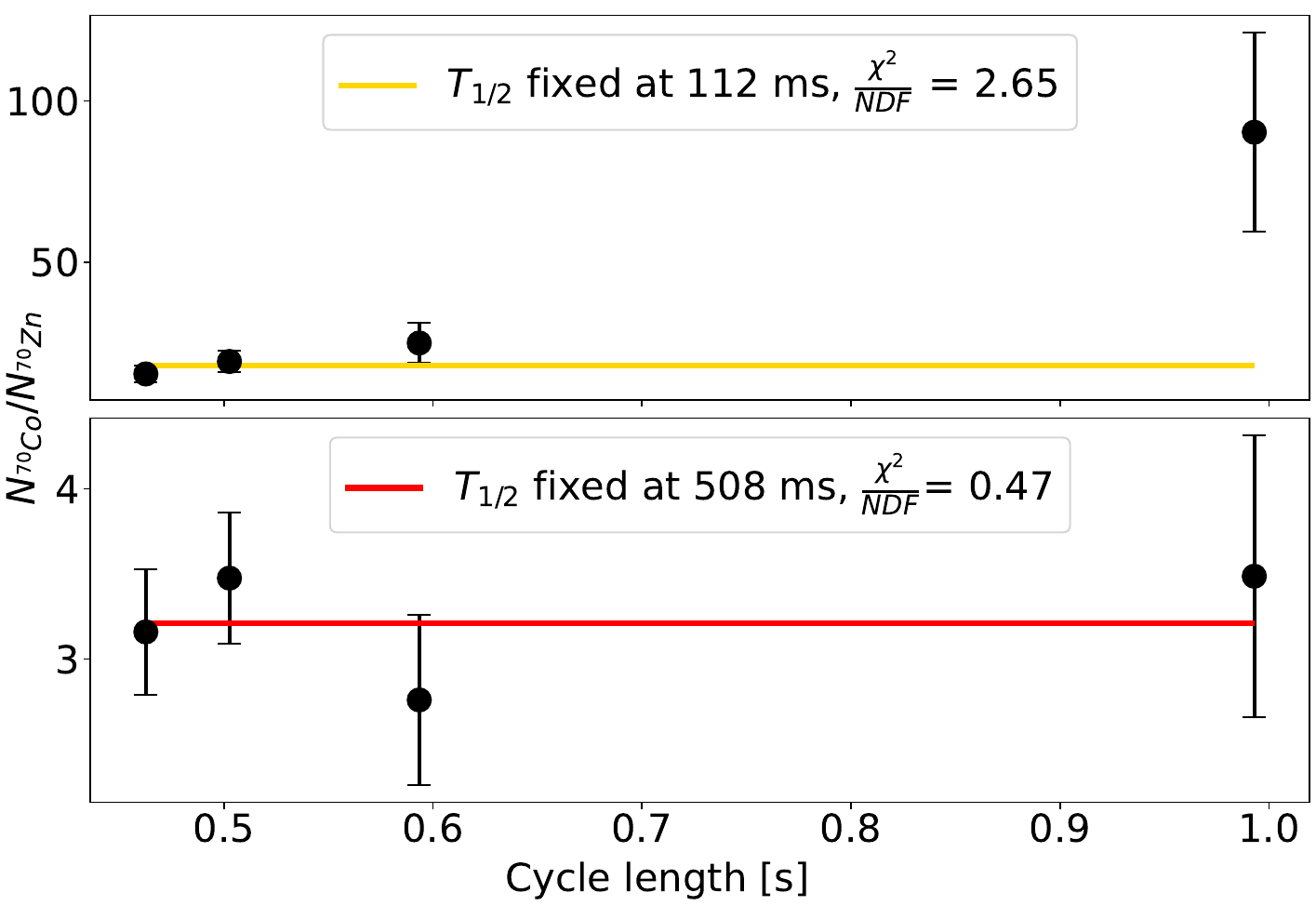}
\caption[]{\justifying Ratio of the initial $^{70}$Co ground-state rate to the initial $^{70}$Zn rate as a function of the cycle length. The $^{70}$Co ground-state rate is calculated using the two experimentally known half-life values (112 ms for the upper panel and 508 ms for the lower panel) from \cite{siegl__2024}. The red and yellow curves correspond to non-linear least-squares fits assuming a constant value in both scenarios.}
\label{Decay70}
\end{figure}

In addition to the mass measurements, the ordering of the two long-lived states in $^{68}$Co and $^{70}$Co was investigated. In previous $\beta-$decay experiments \cite{mueller_beta_1999, flavigny_characterization_2015, morales_type_2017, siegl__2024}, two states had been identified for each isotope, with tentative spin-parity assignments and half-life measurements, but the ground vs. isomeric state ordering could not be established.
Using the PI-ICR technique, the two states can be clearly resolved. By varying the accumulation time ($T_{acc}$) and counting ions, one can determine which of the two half-lives provides the best agreement with the data. 
Count rates were normalized to a stable reference ion present in the beam, to account for beam fluctuations.
Since $T_{acc}$ was not the only parameter modified in the different measurements, the relative rates before any accumulation in the cooler-buncher and the traps (and thus expected to be constant) were computed assuming one or the other possible half-lives, as illustrated for $^{70}$Co in Figure~\ref{Decay70}. A reduced chi-square analysis of the fits provides strong statistical support for the state assignments ($>2\sigma$). For $^{70}$Co, the long-lived low-spin state has clearly been identified as the ground state, whereas for $^{68}$Co, the data strongly favors the short-lived high-spin state as the ground state.\\

\textit{Discussion} - In the following, the deduced spectroscopy is interpreted using Large Scale Shell-Model (LSSM)~\cite{caurier_shell_2005} diagonalizations. The valence space is composed of the $pf$ shell for protons and the $1p_{3/2}$, $0f_{5/2}$, $1p_{1/2}$, $0g_{9/2}$, and $1d_{5/2}$ orbitals for neutrons, on top of an inert $^{48}$Ca core. The Hamiltonian is the LNPS effective interaction~\cite{Lenzi2010} which has been very successful in the studied region~\cite{Rocchini2023,Gade2025}. In this work, new experimental constraints on proton core excitations in $^{73}$Co 
%toward $N=50$ and $^{78}$Ni 
have been adopted~\cite{Co75OakRidge}. 
The remaining physics dealing with the Island of Inversion is left unchanged.

\begin{figure*}
  \centering
    \includegraphics[width=\linewidth]{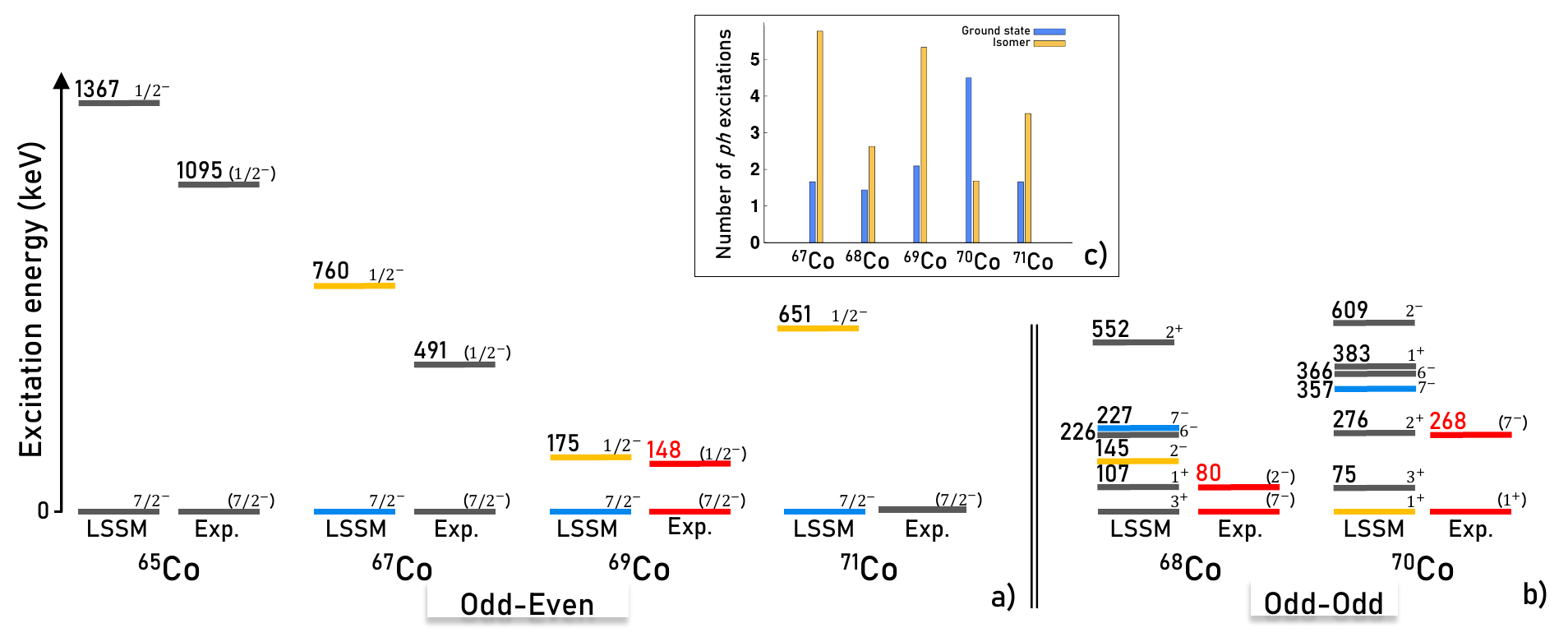}
    \caption[]{\justifying Energy spectra for $^{65,67-71}$Co calculated with LSSM and compared to experimental data, (a) for odd-even nuclei and (b) for odd-odd ones. The levels assigned and measured in this work are shown in red. (c) Calculated number of particle-hole excitations for the ground and isomeric states, shown in blue and orange, respectively. }
    \label{LSSMSpec}
\end{figure*}
Figure~\ref{LSSMSpec} shows the comparison between the experimental low-lying spectra of $^{65,67-71}$Co and theoretical calculations. For the even-$N$ cobalt isotopes (see the left panel of Figure~\ref{LSSMSpec}) , the agreement is very good, showing a rapid lowering of the first excited $1/2^-$ state, reaching a minimum in $^{69}$Co. This energy is expected to increase again in $^{71}$Co. The configurations of these states, in terms of the number of particle-hole excitations across $Z=28$ and $N=40$, are shown in the inset of Figure~\ref{LSSMSpec}. Details of proton and neutron excitations above $Z=28$ and $N=40$, respectively, are given in Table 1 of the Supplemental Material. One can observe a normal structure for the $7/2^-$ ground state and an intruder configuration for the $1/2^-$ isomeric state in $^{67,69,71}$Co.

\begin{figure}[h]
  \centering
  \includegraphics[width=1.0\linewidth]{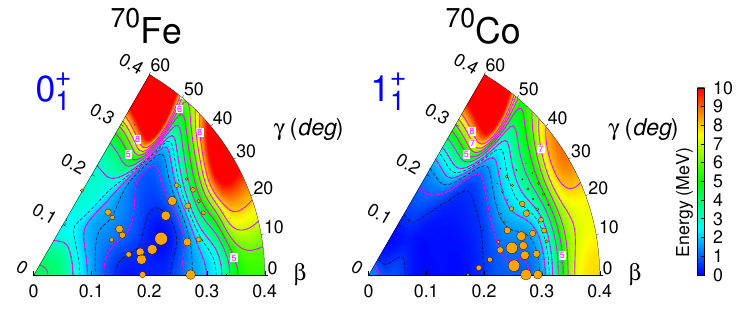}
% \captionsetup{justification=justified} 
  \caption[]{\justifying Parentage of the ground-state and intruder structures between the even-even deformed $^{70}$Fe and the odd-odd $^{70}$Co following the wave-function decomposition onto the quadrupole $(\beta,\gamma)$ deformation components. The orange circles represent the probability of each deformed configuration into the correlated wavefunction.}
  \label{Fe70-Co70}  
\end{figure}

For the odd-odd cobalt isotopes (see the right panel of Figure~\ref{LSSMSpec}), the situation is more complex, with competing positive- and negative-parity states within a narrow energy range. Nevertheless, the reproduction is satisfactory, particularly for $^{70}$Co, where a positive-parity state becomes the ground state, with a high-spin negative-parity state lying at a few hundreds of keV. The configuration of the ground state, shown in the inset of Figure~\ref{LSSMSpec}, corresponds to an intruder configuration. This inversion is only present in $^{70}$Co and appears at the transition between the deformation regime in iron isotopes and the normal regime in nickel isotopes. The present experimental ground-state assignment and its associated theoretical interpretation are at variance with another recent beta-decay study~\cite{Dembski2025acg}, but are consistent with the scenario proposed by Morales et al. in their beta-decay study of $A=70$ nuclei~\cite{morales_type_2017}.\\
\indent For additional insight into shape coexistence, we perform Discrete Non-Orthogonal Shell-Model (DNO-SM) calculations. With this recent development~\cite{dao_nuclear_2022,dao_nuclear_2025}, we diagonalize the same effective interaction within the same valence space as in LSSM in a basis set of the most relevant deformed Hartree-Fock states defined by $\beta$ (axial deformation) and $\gamma$ (triaxial) degrees of freedom (see \cite{dao_nuclear_2022} for details of this expansion). Figure S1 of the Supplemental Material shows the resulting Potential Energy Surfaces (PES) and expansions of the wave-functions. 
For each even-$N$ cobalt isotope, we compute the $7/2^-$ ground state and the first excited $1/2^-$ intruder state, whereas for each corresponding nickel isotope $(Z+1,N)$, we compute the ground state and the excited proton-intruder $0^+$ state. The cobalt and corresponding nickel ground states all exhibit spherical components and similar PES patterns, whereas the excited states are dominated by strongly deformed configurations. This similarity reflects the coupling of a proton hole in cobalt to a prolate-deformed nickel core.\\
\indent We discuss now the structure of the 1$^+$ ground state of $^{70}$Co. Figure~\ref{Fe70-Co70} shows the PES and the expansion of the wave-functions in the PES for both $^{70}$Fe and $^{70}$Co ground states. 
The collective character of $^{70}$Fe, evidenced by its low-lying 2$^+$ state at 480(13) keV~\cite{Santamaria-PhysRevLett.115.192501}, is well reproduced 
in the calculations, showing sizable deformed components with $\beta$= 0.2-0.3.
A similar pattern is found for $^{70}$Co, which is one proton particle and one neutron hole coupled to $^{70}$Fe, supporting its deformed character.

\begin{figure}[h]
   \includegraphics[width=\linewidth]{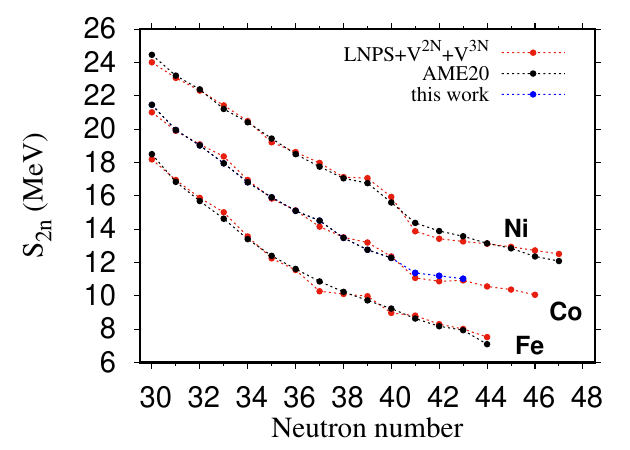}
    \caption[]{\justifying Two-neutron separation energies for iron, cobalt and nickel isotopes, derived from experimental mass values, shown in black for the data from literature~\cite{huang_ame_2021, porter_mapping_2022}, and in blue for this work. Calculations with the LNPS+V$^{2N}$+V$^{3N}$ interaction are shown in red. For better clarity, the iron and nickel values have been shifted by $-2$ and $+2$ MeV, respectively.}
    \label{S2ntheo-3N}
\end{figure}
Finally, we calculate the binding energies for the cobalt and the neighboring iron and nickel chains for $28 \leq N \leq 47$. 
The original LNPS interaction was not constrained on binding energies, with the $f_{7/2}$ single-particle energy set to zero relative to $^{40}$Ca.
As shown in~\cite{Abzouzi,Caurier99-PhysRevC.59.2033}, within the shell-model, binding energies can be determined independently of the spectroscopy by adding to the Hamiltonian terms that depend only on the total number of valence protons and neutrons. 
In the present work, we supplement the LNPS Hamiltonian with such a "master" monopole two-body (V$^{2N}$) or two- plus three-body (V$^{2N}$+V$^{3N}$) term.
Details are given in the Supplemental Material, where the deviations between experimental and shell-model binding energies are also shown (see Fig.~S2). This comparison yields rms deviations of only
$0.514$~MeV and $0.214$~MeV when using quadratic and cubic master terms, respectively. The inclusion of the three-nucleon term thus significantly improves the reproduction of the experimental masses. 
The origin of such a correction is consistent with the findings of ab-initio theory on the need for three-body forces for mass description~\cite{O28-PhysRevLett.105.032501} but also with Many-Body Perturbation Theory derivations of valence-space effective interactions~\cite{PovesZuker81,Napoli2012}.

Figure~\ref{S2ntheo-3N} shows the $S_{2n}$ values for the iron, cobalt and nickel isotopic chains, derived from experimental mass values and compared with LSSM calculations using the $V^{2N}$+$V^{3N}$ correction (see Fig.~S4 in the Supplemental Material for $V^{2N}$ only). The excellent reproduction of binding energies with the $V^{2N}$+$V^{3N}$ master term translates into a very good description of the experimental $S_{2n}$ trends. 
As the cobalt isotopes can be viewed as a single proton hole on top of spherical nickel isotopes, we observe a rather similar trend in $S_{2n}$ for both chains, up to $N=41$. The small structures arise mainly from the monopole part of the interaction, and thus from the progressive filling of the orbitals. In contrast, the iron chain, characterized by a gradual increase of collectivity, shows a more regular evolution~\cite{porter_mapping_2022}. 
From $N=41$, both cobalt and nickel exhibit increased shape coexistence, where the ground states contain a sizable amount of particle-hole excitations. As a result, a steep rise is observed in the multipole part of the $S_{2n}$ in both chains (Fig. S3 of the Supp. Mat.), which becomes even larger at the intruder inversion in $^{70}$Co. 
Finally, the delicate balance between mean-field evolution and pairing + quadrupole correlations nearly equilibrates, producing almost a plateau in the $S_{2n}$ evolution of the cobalt isotopes. Moving further towards $N=50$, the intruder correlations are hindered by Pauli blocking and the trend recovers a smooth decrease, similar to the nickel chain. In conclusion, the $S_{2n}$ evolution showing a trend change in cobalt isotopes can be understood as a fingerprint of the shape coexistence and intruder mixing discussed earlier in the spectroscopy beyond $N\sim 40$. \\

In summary, mass measurements of $^{68-70}$Co were performed using the JYFLTRAP setup at the IGISOL facility. The isomeric and ground states could be separated, their binding energies individually measured and the ordering of the states established. These results clarified the mass surface in this region, showing an excellent agreement of the derived $S_{2n}$ values with LSSM calculations. Finally, an inversion between the intruder and the near-spherical state was revealed in $^{70}$Co. Although this inversion appears to be singular, with a predicted $1/2^-$ intruder state in $^{71}$Co at 650 keV, these findings highlight the need for further mass measurements toward more neutron-rich cobalt isotopes.

\begin{acknowledgments}
The authors would like to thank the mobility support from Projet International de Coopération Scientifique Manipulation of Ions in Traps and Ion sourCes for Atomic and Nuclear Spectroscopy (MITICANS) of CNRS. D.D.D. and F.N. acknowledge  the financial support from CNRS/IN2P3, France, via ABI-CONFI and  ABI-CONFI-II Master projects. D.D.D. acknowledges the financial support from CNRS/IN2P3, France, through the budget for early-career CNRS researcher. A.K. acknowledges the support from the Research Council of Finland under the grants No. 354968, 369714 and 374065 and from the European Union’s Horizon 2020 research and innovation program under Grant Agreement No. 771036 (ERC CoG MAIDEN).
\end{acknowledgments}
\bibliography{ref}

\end{document}

\preprint{APS/123-QED}

\title{Supplemental Material:  Island of Inversion in neutron-rich cobalt isotopes revealed from mass measurements }

\maketitle

%%%%% List of content
%%% - Plot comparison of selecting 958-keV gamma ray and combination of the three transitions 
%%% - Angular distribution plots
%%% - Shell model calculations 
%%% - Kinematic correction of the protons with Geant4 simulation 

%%%%%%%%%%%%%%%%%%%%%%%%%%%%%%%%%%%%%%%%%%%%%%%%%%
This Supplemental Material contains further details of the various experimental and theoretical approaches used in this work.

\section{Experiment: beam production and preparation}

The nuclei of interest were produced through fission induced by 35-MeV protons impinging into a 15 mg/cm2-thick uranium-molybdenum alloy target. The fission fragments were stopped and thermalized in a helium-filled gas cell and extracted by a gas flow and electric fields via a sextupole ion guide~\cite{karvonen_sextupole_2008}. 
The ions were accelerated to 30 qkeV and transported through a 55$^\circ$ dipole magnet with an approximate mass resolving power $M/\Delta M$ of 500. This step allowed for mass separation, ensuring that only ions with similar mass-to-charge ratios (A/q) to the ions of interest (IoI) were selected. The ions were then directed to a radiofrequency quadrupole (RFQ) cooler-buncher~\cite{nieminen_beam_2001} and extracted with the recently commissioned mini-buncher \cite{VIRTANEN2025170186}, before being sent to the JYFLTRAP mass spectrometer~\cite{eronen_jyfltrap_2012}.

This spectrometer consists of two cylindrical Penning traps operating in a 7 T magnetic field. The first trap, filled with helium buffer gas, is used to perform isobaric purification using the buffer-gas cooling technique~\cite{savard_new_1991}. It also provides an $A/q$ separation, being usually 200 to 600 times greater than the one of the dipole magnet. After this cleaning step, the ions of interest were transferred to the second trap for the precise mass measurements using the PI-ICR technique~\cite{eliseev_phase-imaging_2013, nesterenko_phase-imaging_2018}.
This technique is based on the measurement of the phase evolution of radial ion motions in the Penning trap. The radial position in the trap is projected onto a position-sensitive MCP detector. In this work, the cyclotron frequency of the ions was directly measured, using the time pattern shown in the Fig. 3 of Ref \cite{nesterenko_study_2021}.

\section{Calculations}
\subsection{Large-Scale Shell Model (LSSM) and Discrete Non-Orthogonal Shell Model (DNO-SM)}
\begin{center}
\begin{figure}[hb!]
  \centering
  \includegraphics[scale=0.7]{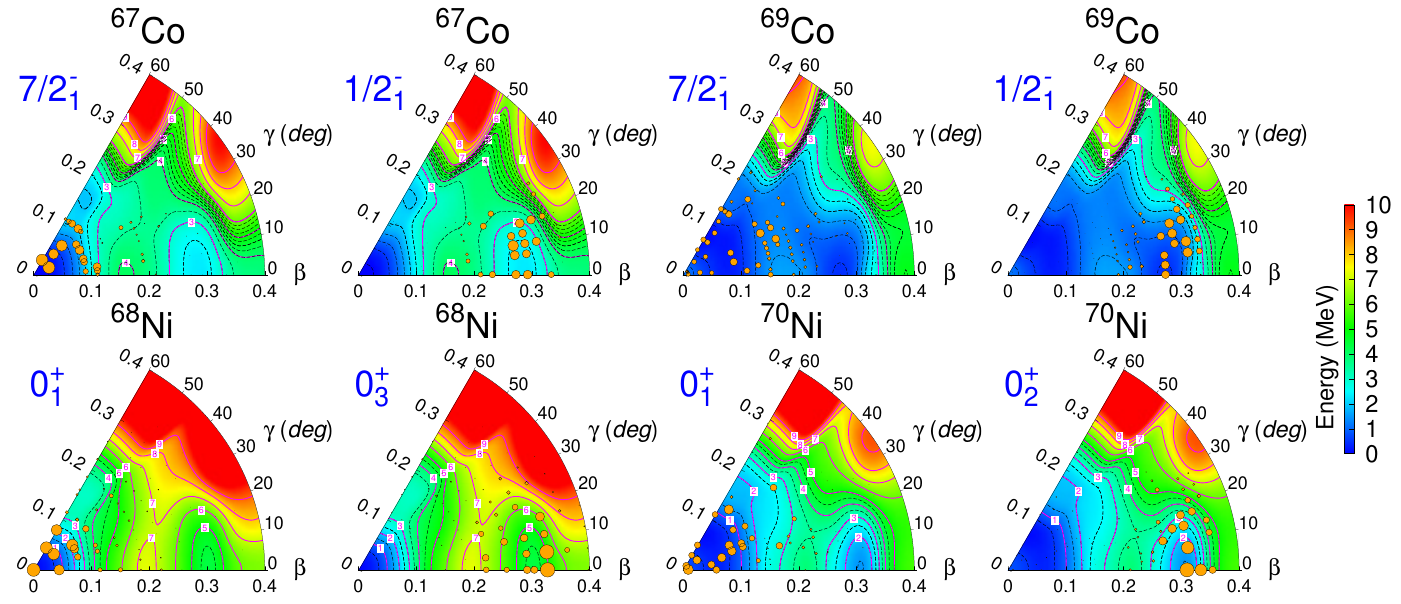}
  \caption{ \justifying Parentage of the ground-state and intruder structures between odd-mass $^{67,69}$Co and even-even $^{68,70}$Ni following the wave-function decomposition onto the quadrupole $(\beta,\gamma)$ deformation components.
   The orange circles represent the probability of each deformed configuration into the correlated wavefunction.}
  \label{Co67-69}
\end{figure}    
\end{center}

For the exact LSSM calculations, shell-model diagonalization was performed using the ANTOINE code~\cite{Antoine1,Antoine2} allowing for up to 13p-13h excitations across the $Z=28$, $N=40$ shell gaps for most of the nuclei. The low-lying sprectrum is  computed for each nucleus of interest and we extract in Table~\ref{OccupationsSG} the orbital occupancies in the valence space.
The DNO shell model~\cite{DNO,nkck-ctvd,5jh8-tdc6} is an alternative tool to solve the secular shell-model eigenvalue problem in the same valence space and interaction, but allowing larger configurational spaces. It has been recently developed to a variation-after-projection (VAP) extension on angular momemtum. The VAP approach was introduced in Refs.~\cite{Schmid1984A_PhysRevC.29.291,Schmid1984B_PhysRevC.29.308,Schmid2004} and more recently in~\cite{QVSM-PhysRevC.103.014312}. In our case, the nuclear states are parametrized in terms of intrinsic non-orthogonal Slater determinants which, while preserving the particle numbers, break the rotational and parity symmetries. Both symmetries are restored consistently, with the use of angular momentum projection techniques, by the minimization of the projected energy. These double variation calculations thus directly determine the involved intrinsic Slater states and their corresponding mixing amplitudes. For a more detailed presentation, we refer to our recent work in Ref.~\cite{dao_nuclear_2025}. \\
In Fig.~\ref{Co67-69}, we show the DNO-SM calculations for the $\frac{7}{2}^-$ ground-state and $\frac{1}{2}^-$ intruder state in odd-mass $^{67-69}$Co as well as the corresponding $0^+_{1,2}$ ground and intruder states in $^{68-70}$Ni. 
For each case, the wavefunction expansion in the Potential Energy Surface is shown, with orange circles indicating the various components of the state in the $(\beta,\gamma)$ plane. 

\subsection{Binding energy calculations}
To evaluate binding energies, we start from the monopole master term extraction from the Duflo-Zuker microscopic mass formula~\cite{DZmass94,AnatomyDZ}: a general 
shell-model Hamiltonian can be split into its monopole and multipole parts $H = H_m + H_M$ where $H_m$ is reponsible for saturation properties and shell evolution, and $H_M$ carries the correlations (pairing, quadrupole, octupole ...).
The  general form of the monopole Hamiltonian contains all scalar particle number operators $\hat n_i$ and can be written as
\begin{equation}
\begin{aligned}
H_m =  \sum_i \varepsilon_i.\hat n_i + \sum_{ij}V_{ij}\hat n_i.\hat n_j \\ + \sum_{ijk}V_{ijk}\hat n_i\hat n_j \hat n_k 
\label{eq:DZmonop}
\end{aligned}
\end{equation}
where $i,j,k$ label proton and neutron single particle orbitals, and $V_{ij}/V_{ijk}$ are the centroids of the interaction expressed as linear combinations of the interaction matrix elements (see II.B in~\cite{RevModPhys.77.427}).
%\begin{equation}
%\begin{aligned}
%V_{ij} = \frac{\sum_J (2J+1)V_{ijij}^J}{\sum_J (2J+1)} 
%\label{eq:centroids}
%\end{aligned}
%\end{equation}
One can then rewrite~\eqref{eq:DZmonop} to extract one subset of two-body (three-body)  matrix elements (TBME) responsible for the masses only, without altering the spectroscopy~\cite{Abzouzi}. In practice, this amounts to extract the same quantity from each of the individual centroids. This procedure has been used in Ref.~\cite{Caurier99-PhysRevC.59.2033} and contains two and two+three body contributions for like-particles and proton-neutron interactions:
\begin{equation}
\begin{aligned}
H_m^{master} = \varepsilon_\nu.n + \varepsilon_\pi.z  \\ \textcolor{black}{+ V^{2N}_{\nu\nu}.\frac{n(n-1)}{2} + V^{2N}_{\pi\pi}.\frac{z(z-1)}{2} +  
 V^{2N}_{\nu\pi}.nz}  \\
 \textcolor{black}{+ V^{3N}_{\nu\nu\nu}.\frac{n(n-1)(n-2)}{6} + V^{3N}_{\pi\pi\pi}.\frac{z(z-1)(z-2)}{6}}  \\
  \textcolor{black}{+ V^{3N}_{\nu\nu\pi}.\frac{n(n-1)}{2}z + V^{3N}_{\nu\pi\pi}.n\frac{z(z-1)}{2}}
\label{eq:DZmaster}
\end{aligned}
\end{equation} 
Here $\hat{n},\hat z= \sum_i \hat n_i $  are the valence neutron and proton numbers respectively, and the various total centroids $V^{2N}_{\nu\nu}$, $V^{2N}_{\pi\pi}$ ... are linear combinations of all the $V_{ij}/V_{ijk}$ centroids defined in~\eqref{eq:DZmonop}~\cite{Abzouzi}.
In the present case, the $^{48}$Ca core is asymmetric and induces distinct proton and neutron parameters, with no charge symmetry retained.

\begin{figure*}[h]
\centering
    \includegraphics[scale=0.65]{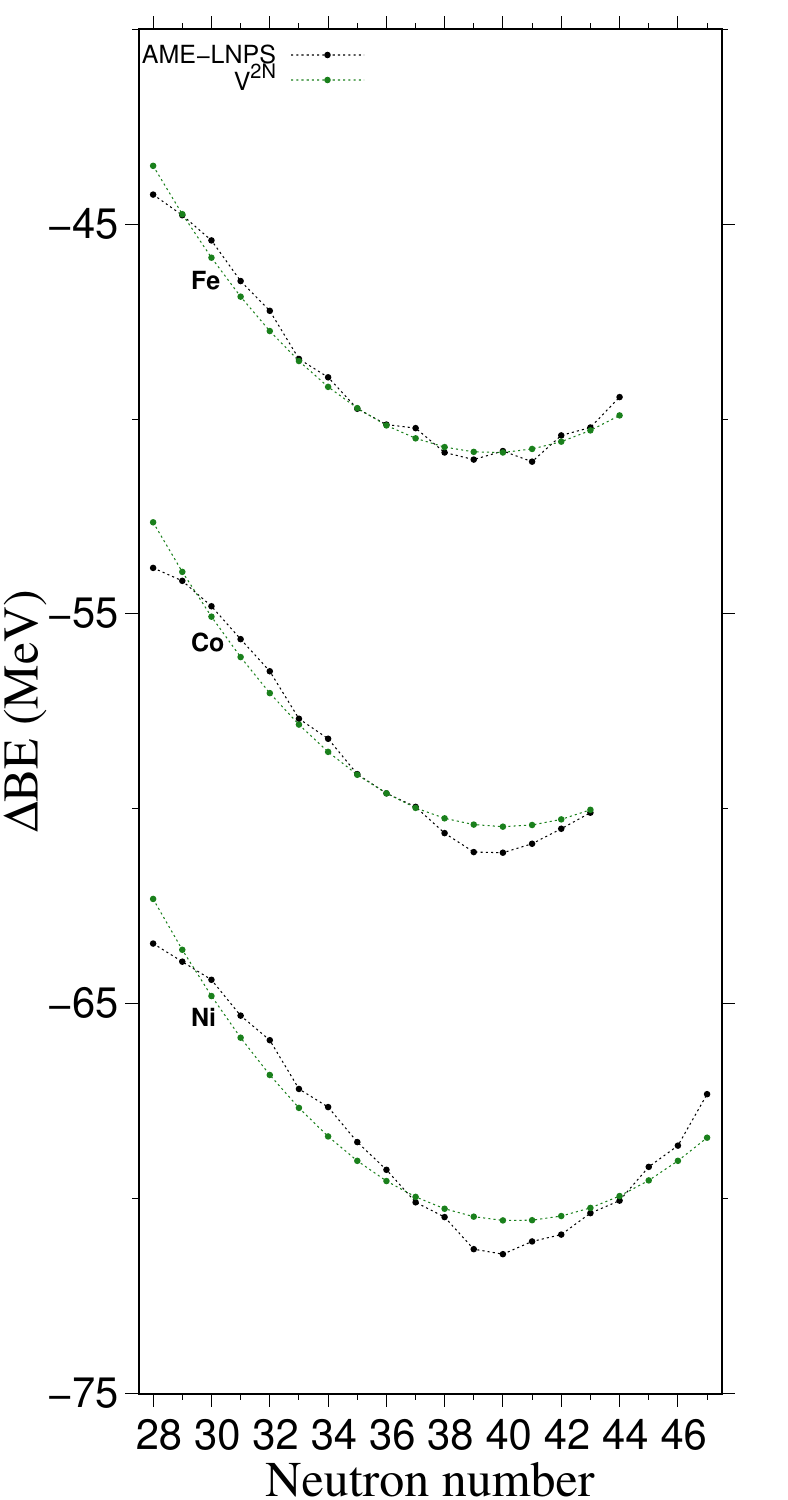}
    \includegraphics[scale=0.65]{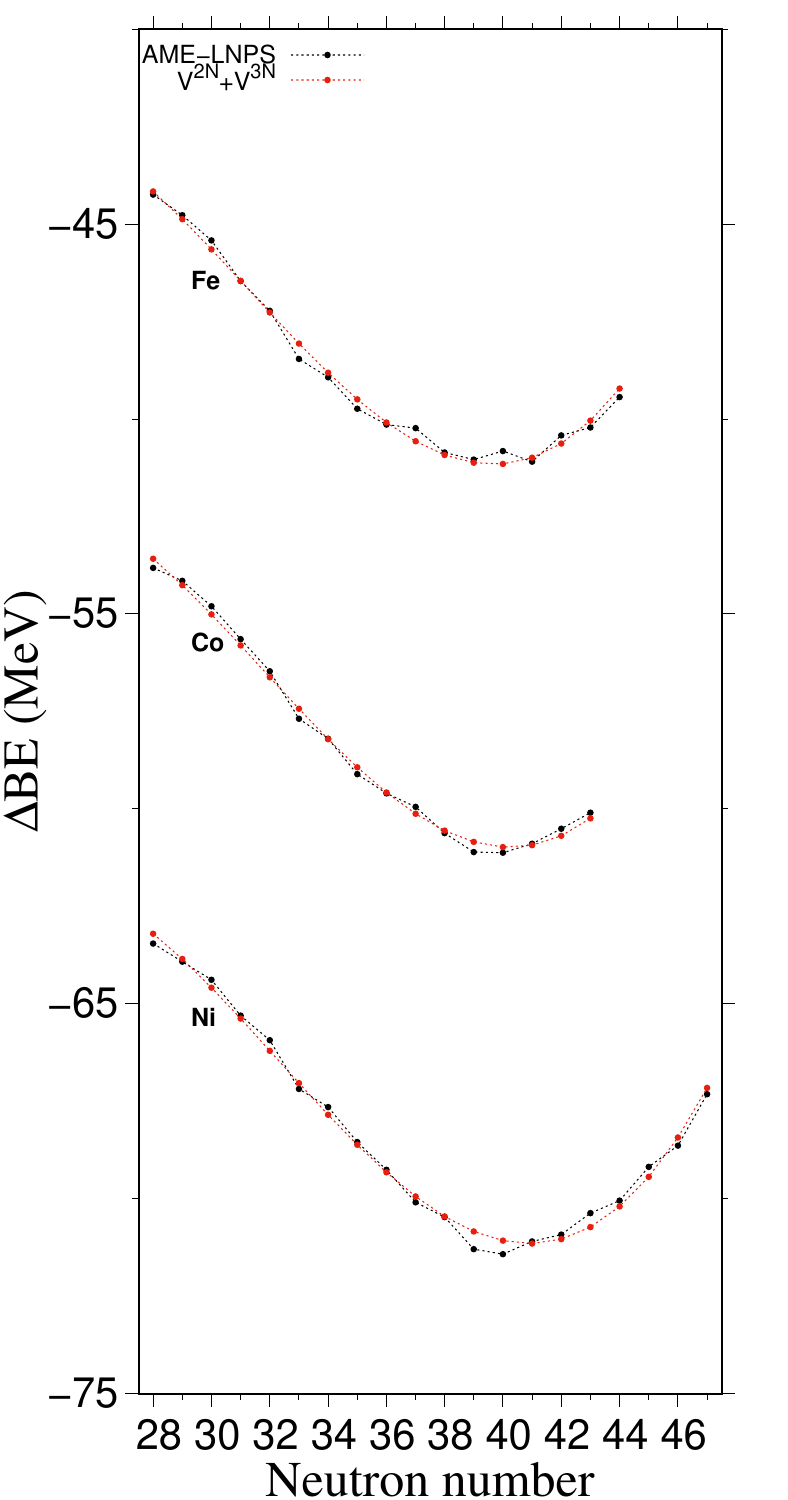}
    \caption{\justifying Difference between experiment and shell model based binding energies $\Delta$BE using two-body (V$^{2N}$) and two-body + three-body (V$^{2N}$+ V$^{3N}$) master terms for Fe, Co and Ni. The experimental mass data is from literature~\cite{huang_ame_2021, porter_mapping_2022}, except the masses from this work.}
    \label{DeltaBE}
\end{figure*}

The Coulomb energies are evaluated using the simple formula for $E_C$ from~\cite{AnatomyDZ}:
\begin{equation}
\begin{aligned}
%E_C& = \frac{Z(Z-1)+0.76[Z(Z-1)^{2/3}]}{r_c}  \\
% \mathrm{with}\quad  r_c& =A^{1/3}\Bigg[1-\bigg(\frac{T}{A}\bigg)^2\Bigg]
E_C&= +0.7\frac{Z^2}{A^{1/3}}
\end{aligned}
\end{equation}
where $Z$ and $A$ are the total proton and mass numbers, respectively.\\ 
Fig.~\ref{DeltaBE} shows the results of the fit of the residuals between the LNPS initial calculations and the experimental data from AME and data from this work. 
The quadratic (left) and cubic (right) master terms from Eq.~\ref{eq:DZmaster} are shown. 
After calculations of the total binding energies for iron, cobalt and nickel chains, the two-neutron separation energies are shown in Figure~\ref{S2ntheo-2N} and in Figure~5 of the main article for the V$^{2N}$ and V$^{2N}+$V$^{3N}$ master terms respectively.

\begin{figure}[h]
\centering
     \includegraphics[trim=0 0 0 0 , clip,width=1.0\columnwidth]{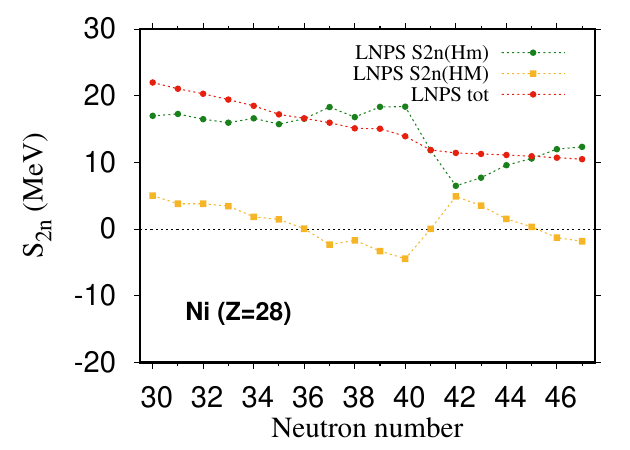}
    \includegraphics[trim=0 0 0 0 , clip,width=1.0\columnwidth]{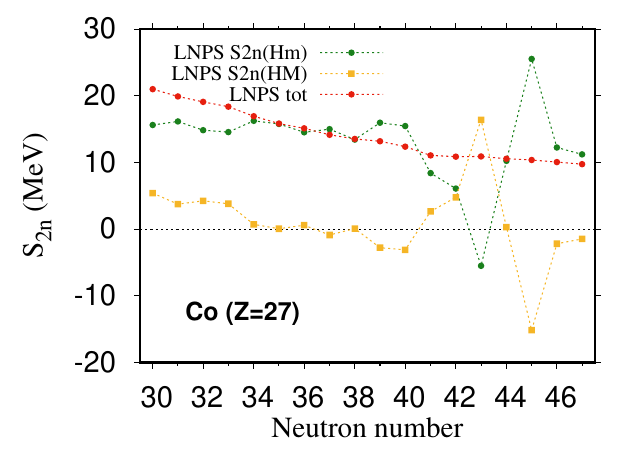}
    \includegraphics[trim=0 0 0 0 , clip,width=1.0\columnwidth]{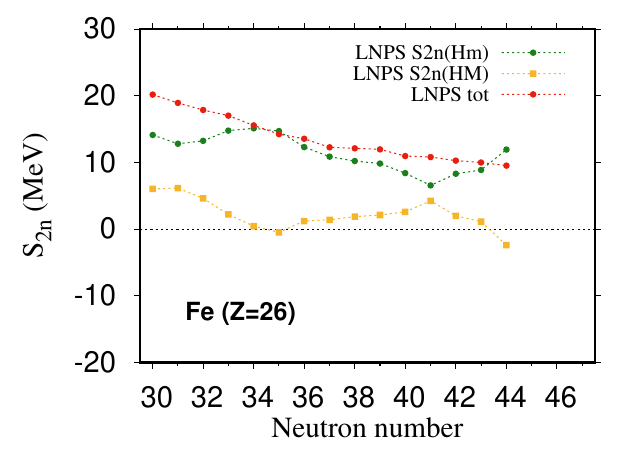}    \caption{\justifying $S_{2n}$ evolution of  the total, monopole (Hm) and multipole hamiltonian (HM) for the nickel (top), cobalt (middle)  and iron (bottom) isotopes.}
    \label{HmHM}
\end{figure}

\begin{figure}[h]
   \includegraphics[trim=0 0 0 0 , clip,width=1.0\columnwidth]{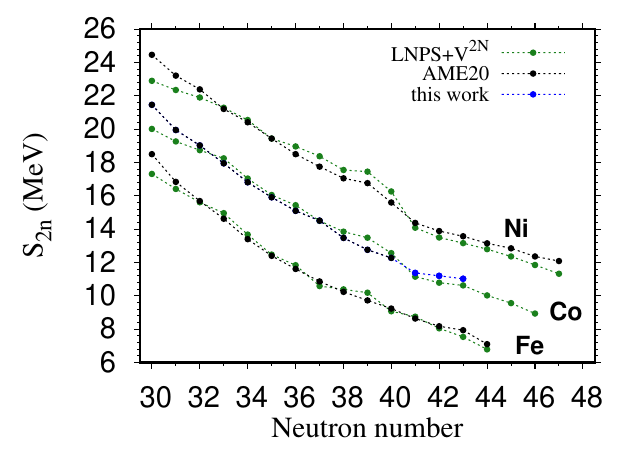}
    \caption{\justifying Two-neutron separation energies for iron, cobalt and nickel isotopes, derived from experimental mass values, shown in black for the data from literature~\cite{huang_ame_2021, porter_mapping_2022}, and in blue for this work, and calculated with the LNPS+V$^{2N}$ interaction (green). For better clarity, the iron and nickel values have been shifted by $-2$ and $+2$ MeV, respectively.}
    \label{S2ntheo-2N}
\end{figure}

Finally, for better insight into the physics at play, we decompose the total Hamiltonian into its monopole part $H_m$, which carries the mean-field evolution, and its multipole part $H_M$, which carries the correlations. For each state, we evaluate their relative contributions to the total binding energies and plot their $S_{2N}$ evolution in Figure~\ref{HmHM} for each isotopic chain.

\onecolumngrid
\twocolumngrid

\begin{table*}[h!]
\centering
%\scalebox{0.9}{      
       \begin{tabular}{llll cclc cccc}
          \hline\hline
                   &         & E*$_{exp}$ & E*$_{theo}$ &  $1f_\frac{7}{2}$       & $2p_\frac{3}{2}$ & $1f_\frac{5}{2}$ & $2p_\frac{1}{2}$ & $1g_\frac{9}{2}$ & $2d_\frac{5}{2}$ & $\Delta n^\pi(p_\frac{3}{2}+f_\frac{5}{2}+p_\frac{1}{2})$ & $\Delta n^\nu(g_\frac{9}{2}+d_\frac{5}{2})$\\
& Nuclide& $I^{\pi}$ & &  & & & & & & &  \\[5pt]
            \hline 
%$^{65}$Co  & $7/2^- $ & 0.000 & 0.000 &   $6.46$ & $0.33$ & $0.18$ & $0.026$ &        &        & \\
%           &          &   &       &   $8.00$ & $3.76$ & $4.07$ & $1.12$ & $0.98$ & $0.07$ & 1.05 \\
%           & $1/2^-$  & 1.095 & 1.367 & $5.36 $ & $0.82$ & $0.63$ & $0.20$ &        &   &     \\
%           &          & &      &  $8.00$ & $3.59$ & $3.32$ & $0.63$ & $2.20$ & $0.26$ & {\bf 2.46} \\[5pt]
$^{67}$Co  & $7/2^- $ & 0.000 & 0.000 &  $6.65$ & $0.19$ & $0.14$ & $0.019$ &        &   &   $0.35$  & \\
           &          & &      &   $8.00$ & $3.89$ & $5.22$ & $1.58$ & $1.22$ & $0.09$ & & 1.32\\
           & $1/2^-$  & 0.492  & 0.760  & $5.25 $ & $0.74$ & $0.77$ & $0.24$ &        & &    {\bf 1.77} &  \\
           &          & &       &   $8.00$ & $3.79$ & $3.65$ & $0.54$ & $3.47$ & $0.55$ & &{\bf 4.02} \\[5pt]
$^{68}$Co  & $7^- $ & 0.000 & 0.227 &  $6.65$ & $0.19$ & $0.14$ & $0.019$ &        &   &  0.35   \\
           &          & &      &   $8.00$ & $3.91$ & $5.40$ & $1.62$ & $1.97$ & $0.091$ & & 1.06 \\
           & $2^-$  & 0.080 & 0.145 & $5.21 $ & $0.76$ & $0.79$ & $0.24$ &        &  &  {\bf  1.79} &  \\
           &          & &      &   $8.00$ & $3.85$ & $3.71$ & $0.46$ & $4.35$ & $0.62$ & &{\bf  3.97} \\[5pt]
$^{69}$Co  & $7/2^- $ & 0.000 & 0.000 &  $6.44$ & $0.32$ & $0.21$ & $0.034$ &        &   &  0.58   \\
           &          & &      &   $8.00$ & $3.89$ & $5.17$ & $1.37$ & $3.28$ & $0.28$ & &1.54 \\
           & $1/2^-$  & 0.148 & 0.175 & $5.23 $ & $0.71$ & $0.82$ & $0.24$ &        &  &  {\bf  1.77} &   \\
           &          & &      &   $8.00$ & $3.85$ & $3.93$ & $0.66$ & $4.81$ & $0.74$ & &{\bf 3.55} \\[5pt]
$^{70}$Co  & $1^+ $ & 0.000 & 0.000 &  $5.29$ & $0.65$ & $0.83$ & $0.23$ &        &   &  {\bf 1.71} &  \\
           &          & &      &   $8.00$ & $3.93$ & $4.14$ & $1.14$ & $5.13$ & $0.66$ & &{\bf 2.79} \\
           & $7^-$  & 0.268 & 0.357 & $6.49 $ & $0.29$ & $0.19$ & $0.029$ &        &  &  0.51    \\
           &          & &      &   $8.00$ & $3.92$ & $5.46$ & $1.44$ & $3.90$ & $0.27$ & & 1.17 \\[5pt]
$^{71}$Co  & $7/2^- $ & 0.000 & 0.000 &   $6.45$ & $0.31$ & $0.21$ & $0.031$ &        &        & 0.55\\
           &          & &      &   $8.00$ & $3.93$ & $5.51$ & $1.45$ & $4.69$ & $0.42$ & & 1.11 \\
           & $1/2^-$  & - & 0.651 & $5.36 $ & $0.61$ & $0.80$ & $0.23$ &        &        & {\bf 1.64} & \\
           &          & &      &   $8.00$ & $3.92$ & $4.84$ & $1.34$ & $5.20$ & $0.69$ & & {\bf 1.89} \\[5pt]
%$^{73}$Co  & $7/2^- $ & 0.000& 0.000 &   $6.54$ & $0.25$ & $0.19$ & $0.023$ &        &        & \\
%           &          & &      &   $8.00$ & $3.97$ & $5.82$ & $1.76$ & $5.96$ & $0.50$ & 0.46 \\
%           & $1/2^-$  & 1.178 & 1.019 &  $5.50 $ & $0.50$ & $0.78$ & $0.22$ &        &    &    \\
%           &          & &      &   & $3.97$ & $5.80$ & $1.83$ & $5.67$ & $0.73$ & 0.40 \\[5pt]
%$^{75}$Co  & $7/2^- $ & 0.000 & 0.000 &     $6.65$ & $0.16$ & $0.17$ & $0.016$ &        &        & \\
%           &          & &      &     $8.00$ & $3.98$ & $5.93$ & $1.87$ & $7.82$ & $0.40$ & 0.22 \\
%           & $1/2^-$  & 1.914 & 1.929 &  $5.54$ & $0.40$ & $0.84$ & $0.21$ &        &    &    \\
%           &          & &      &    $8.00$ & $3.98$ & $5.93$ & $1.92$ & $7.45$ & $0.71$ & 0.16 \\           
    \hline\hline
      \end{tabular}           
\caption{\justifying Occupations of valence-shell orbitals in LSSM calculations for ground state and isomeric states in $^{67-71}$Co. For each state, proton and neutron occupations are given in the first and second line, respectively. $\Delta n ^{\pi}$ correspond to the number of proton excitations to the upper $pf$ orbitals, whereas $\Delta n ^{\nu}$ correspond to the number of neutron excitations to the upper $gd$ orbitals. Intruder structures are indicated in bold. Experimental and theoretical excitation energies, E*$_{exp}$ and E*$_{theo}$, are given in MeV.}
\label{OccupationsSG}
\end{table*}
%\end{center}
\twocolumngrid
\clearpage
\bibliographystyle{apsrev4-2.bst}
\bibliography{ref.bib}
%%%%% END OF REGULAR SUPPLEMENTAL MATERIAL